\documentclass[aps,prl,reprint,superscriptaddress]{revtex4-2}
\usepackage{graphicx}
\usepackage{dcolumn}
\usepackage{bm}
\usepackage{amsmath}
\usepackage{comment}
\usepackage{xcolor}

\begin{document}

\title{Fingerprints of unconventional pairing in superconductor-hole-gas heterostructures}
 \author{Serafim S. Babkin}
     \affiliation{Institute of Science and Technology Austria (ISTA), Am Campus 1, 3400 Klosterneuburg, Austria}
    \author{Karsten Flensberg}
     \affiliation{Niels Bohr Institute, University of Copenhagen, DK-2100 Copenhagen, Denmark}
    \author{Jeroen Danon}
    \affiliation{Department of Physics, Norwegian University of Science and Technology, NO-7491 Trondheim, Norway}
    \author{Maksym Serbyn}
    \affiliation{Institute of Science and Technology Austria (ISTA), Am Campus 1, 3400 Klosterneuburg, Austria}

\begin{abstract}
Superconductor and two-dimensional hole gas heterostructures are promising platforms for quantum devices, but their intrinsic properties, e.g., unconventional pairings, are hard to access experimentally. Here, we show that the in-plane-field dependence of the superfluid stiffness and spin susceptibility exhibits nonanalyticities encoding the induced gap structure in the hole gas. Additionally, Fourier harmonics of the stiffness tensor's field-angle dependence, accessible via kinetic-inductance measurements, provide fingerprints of unconventional pairing and cubic Rashba coupling.
\end{abstract}
\maketitle
\noindent{\it Introduction.---}Heterostructures composed of superconducting (SC)
and quasi-two-dimensional semiconducting layers have recently attracted considerable theoretical
and experimental attention as 
a platform for unconventional
superconducting states, including $p$-wave and $d$-wave pairing \cite{Geproxy,pino2025},
as well as topological superconductivity \cite{LeijnseReview,FlensbergReview,AguadoReview}. In addition, they are widely
used in quantum devices such as (Andreev) spin qubits~\cite{Janvier2015,Hays2021,PitaVidal2023} and hybrid circuit quantum
electrodynamic devices~\cite{BurkardReview2020}. Recently, particular interest has focused
on heterostructures in which the semiconductor hosts a two-dimensional
hole gas (in, e.g., germanium)~\cite{Hendrickx2018,Tosato2023,Aggarwal2021,Valentini2024,lakic2024}, since these systems can enable all-electric
spin-qubit control~\cite{ScappucciReview,Fang2023} and form transparent
interfaces with conventional $s$-wave superconductors such as aluminum.

This experimental interest has also motivated theoretical studies
of SC--two-dimensional hole gas (HG) heterostructures. 
Our recent work~\cite{Geproxy} (see also~\cite{pino2025}) derived an effective Hamiltonian for the heavy-hole band, revealing a rich mixture of $s$-, $p$-, and $d$-wave superconducting pairings and computed the resulting density of states and Bogoliubov Fermi surfaces~\cite{Yuan2018,Zhu2021} in the proximitized semiconductor.  At the same time, immediate experimental verifications of the theoretical predictions of Refs.~\cite{Geproxy,pino2025} are hard: the HG is buried beneath the SC making tunneling measurements difficult, while transport measurements are also problematic because the SC acts as a shunt. In this work we identify the superfluid stiffness and spin susceptibility as two distinct and parametrically independent probes of the induced pairing and microscopic details of the proximitized HG.

Recent experiments have probed the superfluid stiffness via kinetic inductance, $L_K\propto 1/\rho_s$, using a resonant microwave circuit \cite{Phan_2022,Babkin_2024,feyrer2026}, enabling sensitive measurements in the presence of magnetic fields. 
Likewise, the spin susceptibility has been measured in two-dimensional semiconducting layers using thermodynamic magnetization techniques~\cite{PhysRevB.67.205407,Shashkin} and in superconductors using NMR Knight-shift measurements~\cite{Hammond-Kelly}. These experimental precedents motivate adapting spin-sensitive probes to measure the total spin susceptibility of SC--HG heterostructures. Quantitative interpretation of this response, however, requires a microscopic theory that disentangles the HG contribution from the superconducting background.

In this paper, we calculate the superfluid-stiffness tensor and spin susceptibility
using the microscopic theory of Ref.~\cite{Geproxy} and identify three
independent signatures of the proximitized HG:
{\it (i)} step-like non-analyticities in both quantities as a function of
in-plane field $B$, with step locations and heights directly encoding appearance or change in topology of Bogoliubov Fermi surfaces; 
{\it (ii)} a finite spin susceptibility at $B=0$~\cite{Rashba2001} arising from the interplay of proximity-induced pairing and Rashba spin-orbit coupling; 
and {\it (iii)} non-zero higher Fourier harmonics $Q_{n\neq 0}$ of the superfluid-stiffness tensor
as a function of field angle $\phi_B$ giving fingerprints of induced
$p$- and $d$-wave pairing components and cubic Rashba spin-orbit coupling.
In contrast, the SC contribution to {\it (i)} is smooth, and it does not contribute to  {\it (ii)}--{\it (iii)} at all.
Beyond the specific SC--HG platform considered here, our results establish thermodynamic response anomalies and their angular harmonics as a general, transport-free protocol for detecting induced unconventional pairing in hybrid superconducting heterostructures.

\begin{figure*}[t]
  \centering
  \includegraphics[width=1.99\columnwidth]{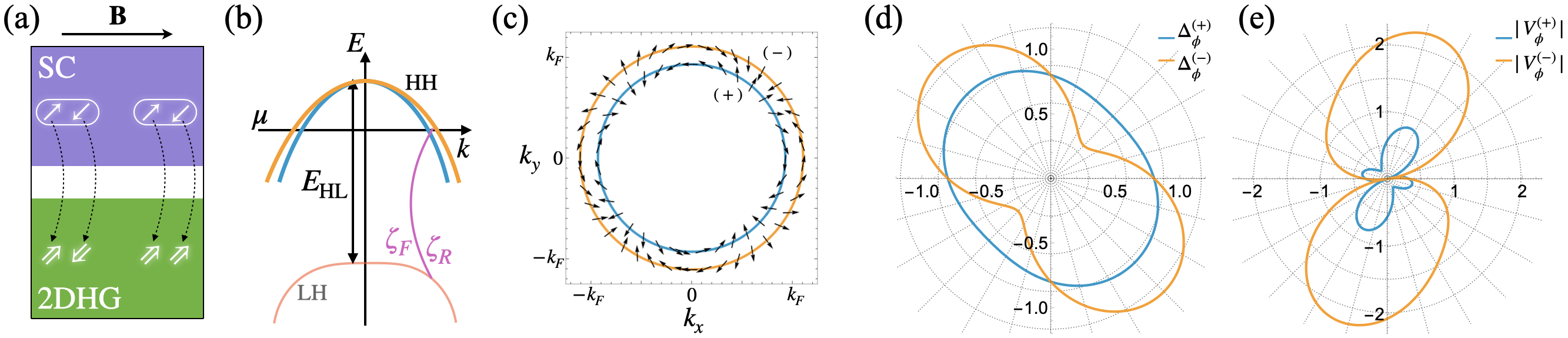}
  \caption{\label{fig1}
  Proximitized HG. (a)~SC--HG heterostructure in an in-plane magnetic field $\mathbf{B}$; interlayer tunneling (dashed arrows) induces superconducting correlations in the hole gas. (b)~Hole band structure: the chemical potential $\mu$ resides in the Rashba-split heavy-hole (HH) bands, separated by $E_{\rm HL}$ from the light-hole (LH) band; $\zeta_{\rm F}$ and $\zeta_{\rm R}$ encode HH--LH mixing. (c)~Fermi surfaces of the Rashba-split bands $(\pm)$, whose spin texture winds three times, reflecting cubic Rashba SOC. (d)-(e)~Interplay of angular dependences of the induced pairing $\Delta_{\phi}^{\pm}$ (d) and Zeeman energy $|V_{\phi}^{\pm}|$ (e) on the Fermi surface produces the angle-dependent gap closures underlying all signatures discussed in the text, both observables are normalized by $\Delta$. The parameters used for the plots are the same as in \cite{Geproxy}: $\gamma_1=13,\ \gamma_s=5,\ \zeta_{\rm F}=0.35,\ E_{\mathrm{HL}}=-100\Delta,\ \alpha_{\rm R} k_F=10\Delta,\ \tau_{+-}=|\tau_{z+}|=0.8,\ \phi_t=\pi/4,\ \zeta_R=-0.2,g=-2,\ \kappa=3.4,\ \phi_B=0,\ \xi_{k_{\rm F},s}=-5\Delta$. For the panel (e), $B=0.4\Delta$, while for others $B=0$.}
\end{figure*}

\noindent{\it Model.---}We introduce the microscopic model for the two-dimensional SC--HG heterostructure, placed in an in-plane magnetic field $\mathbf{B}$, see Fig.~\ref{fig1}(a). The superconductor is assumed to be conventional (spin-singlet $s$-wave
pairing) and is treated within the BCS model, while the HG is modeled by the Luttinger--Kohn Hamiltonian~\cite{Luttinger_Kohn,Luttinger_1956}, see the band structure in Fig.~\ref{fig1}(b). Atomic spin--orbit
coupling (SOC) in the hole gas, together with inversion-symmetry breaking
due to confinement, gives rise to Rashba SOC~\cite{winklerBook}. We consider the
regime, where Zeeman energy is small enough so that Rashba-split bands are well-separated, see Fig.~\ref{fig1}(c). Additionally, we neglect orbital effects from the
magnetic field and retain only Zeeman contributions, which is appropriate for
an in-plane field in a thin film, provided the total thickness of the conducting layer is small compared to the magnetic length.

In the heterostructure, the spatial closeness of the superconducting film and HG [Fig.~\ref{fig1}(a)] enables interlayer tunneling, which
modifies the properties of both layers. The relevant
orbitals are $s$-like in the superconductor and $p$-like in the
HG, and no selection rules were imposed on the hopping, since the interface presumably
breaks rotational symmetry along the $z$-axis.
This hopping strongly renormalizes the HG and, in particular, induces superconducting
correlations in it~\cite{Valentini2024}. In this case, the effective Bogoliubov--de Gennes Hamiltonian for the hybridized HG can be written as \cite{Geproxy}
\begin{align}
\label{H_simp}
 & H=\left(\begin{array}{cc}
H^{-} & 0\\
0 & H^{+}
\end{array}\right), 
\ 
H^{\pm}=\left(\begin{array}{cc}
\Xi_{\mathbf{k}}^{\pm}\mp V_{\phi}^{\pm} & \Delta_{\phi}^{\pm}\\
(\Delta_{\phi}^{\pm})^* & -\Xi_{\mathbf{k}}^{\pm}\mp V_{\phi}^{\pm}
\end{array}\right).
\end{align}
Here the basis is chosen such that the two diagonal 
blocks $H^{\pm}$ describe the two Rashba-split bands, and are determined by the 
band dispersion $\Xi_{k}^{\pm}$, the Zeeman terms $V_{\phi}^{\pm}$, and the effective pairing $\Delta_{\phi}^{\pm}$,
\begin{align}
\Xi_{\mathbf{k}}^{\pm} = {}&{}
\xi_{k} \mp 3\alpha_{\rm R}k\zeta_{\rm F} \nonumber\\ \label{Eq:xi2x2}
& +\xi_{k_{\rm F},s}\tau_{+-} \left(\zeta_{\rm F}\pm\zeta_{\rm R}\right)\cos [2(\phi-\phi_{t})],\\
\nonumber
V_{\phi}^{\pm}={}&{}\frac{1}{2}g B \tau_{+-} \sin\left(\phi_{B}-3\phi+2\phi_{t}\right) \\
\label{Eq:Zeeman2x2}
& -\frac{1}{2}\left(6\kappa + g \tau_{+-}\right)B\left(\zeta_{\rm F}\pm\zeta_{\rm R}\right)\sin\left(\phi_{B}-\phi
\right),\\
\label{Eq:Delta2x2}
 \Delta_{\phi}^{\pm}={}&{}\Delta\tau_{+-}\left[1 - \left(\zeta_{\rm F}\pm\zeta_{\rm R}\right)\cos\left(2\phi-2\phi_{t}\right)\right],
\end{align}
where, $\xi_{{k}}=k^{2}/2m_{\mathrm{HG}}-\mu$ is the kinetic energy and $m_{\mathrm{HG}}$ is the effective mass of electrons in the HG, while $\xi_{k_{\rm F},s}=\xi_{{k},s}|_{k=k_{{\rm F},\mathrm{HG}}}$ is the kinetic energy of electrons in the superconductor evaluated at the Fermi wave vector $k_{F,\mathrm{HG}}$ of the HG. Next, $\alpha_{\rm R}$ is the Rashba SOC strength and $\phi$ is the angle of the in-plane momentum ${\bf k}$. The strength of the proximity effects is parameterized by the amplitude $\tau_{+-}$ and the phase $\phi_{t}$ characterizes details of the interlayer coupling, while $\zeta_{\rm F}$ and $\zeta_{\rm R}$ encode heavy-hole--light-hole mixing (see Ref.~\cite{Geproxy} for an explicit definition of all these quantities). Finally, $\Delta$ is the order parameter in the SC, $\kappa$ is the $g$-factor of electrons in the HG and $g$ is the $g$-factor of electrons in the SC; $B$ and  $\phi_{B}$ denote the magnitude and angle of in-plane magnetic field. 
As we mention before, we work in the regime where the Rashba energy scale is dominant compared to the Zeeman splitting, 
$\kappa B\lesssim\sqrt{\alpha_{\rm R}k_{\rm F}|\xi_{k_{{\rm F}},s}|}$.

The momentum-independent part of $\Delta_{\phi}^{\pm}$ in Eq.~(\ref{Eq:Delta2x2}) corresponds to the induced singlet $s$-wave pairing, while the momentum-dependent part describes unconventional pairing components: singlet $d$-wave, proportional to $\zeta_{\rm F}$, and triplet $p$-wave, proportional to $\zeta_{\rm R}$, resulting in the anisotropic gap illustrated in Fig.~\ref{fig1}(d). Additionally, the first term in $V_{\phi}^{\pm}$ captures the interplay between the Zeeman coupling in the SC and the cubic Rashba SOC in the HG, which is reflected in the $3\phi$ angular dependence, see Fig.~\ref{fig1}(e).

\noindent{\it Observables.---}We calculate experimentally accessible observables for the
heterostructure and identify the signatures of the HG properties. We consider linear
responses of the system to the vector potential $\mathbf{A}$ and
Zeeman field $\mathbf{B}$, and derive the superfluid stiffness $\rho_{s,\alpha\beta}$
and the spin susceptibility $\chi_{\alpha\beta}$, respectively.
We use a field-theoretical formulation: integrating out
fermionic degrees of freedom in the partition function, we
derive the effective action $S\left[\mathbf{A},\mathbf{B}\right]$, whose derivatives give the current density $\mathbf{j}$ and magnetization $\mathbf{M}$, decomposed into contributions from the superconducting layer and the hybridized HG,
\begin{align}
 & j_{\alpha}=\frac{\partial S}{\partial A_{\alpha}},\qquad\frac{1}{4}\frac{\partial j_{\alpha}}{\partial A_{\beta}}=\rho_{s,\alpha\beta}^{\mathrm{SC}}+\rho_{s,\alpha\beta}^{\mathrm{HG}}\\
 & M_{\alpha}=-\frac{\partial S}{\partial B_{\alpha}},\qquad\frac{\partial M_{\alpha}}{\partial B_{\beta}}=\chi_{\alpha\beta}^{\mathrm{SC}}+\chi_{\alpha\beta}^{\mathrm{HG}}.
\end{align}
We emphasize that experiments measure the total responses $\partial j_{\alpha}/\partial A_{\beta}$ and $\partial M_{\alpha}/\partial B_{\beta}$, to which the two layers contribute in parallel.
Below we present the results at \(T=0\); finite temperature broadens the features discussed below but does not lead to qualitatively new behavior (see the Supplemental Material for details).

\noindent{\it Superconducting layer.---}We start with the contributions from the superconducting film.
In the Pauli limit, where pair breaking arises solely from the Zeeman contribution, the superconducting state is suppressed at a critical field $B_P$ via a first-order phase transition~\cite{Clogston1962,Chandrasekhar1962}. Consequently, for $B<B_P$, the spin susceptibility is trivial, $\chi^{\mathrm{SC}}_{\alpha\beta}=0$: the magnetization vanishes because Cooper pairs are spin singlets and the Zeeman energy is insufficient to break them. For the same reason, the superfluid stiffness remains equal to its zero-field value when $B<B_P$:  $\rho_{s,\alpha\beta}^{\mathrm{SC}}=\delta_{\alpha\beta}\rho_{s,0}^{\mathrm{SC}}$, where $\rho_{s,0}^{\mathrm{SC}}=n_{\mathrm{eff}}^{\mathrm{SC}}/(4m_{\mathrm{SC}})$, where $n_{\mathrm{eff}}^{\mathrm{SC}}=k_{{\rm F},\mathrm{SC}}^{2}/(2\pi)\sum_{n_{z}=1}^{n_{z,\mathrm{max}}}[1-\pi^{2}n_{z}^{2}/(d^{2}k_{{\rm F},\mathrm{SC}}^{2})]$ is an effective carrier density. Here,  $k_{F,\mathrm{SC}}$ is Fermi momentum,
$d$ is the film thickness, $n_{z,\mathrm{max}}=\lfloor dk_{{\rm F},\mathrm{SC}}/\pi\rfloor$ is
number of transverse modes and $m_{\mathrm{SC}}$ is the effective electron mass in the superconductor. For $B>B_P$, superconductivity is absent, so the normal-metal results apply: $\rho_{s,\alpha\beta}^{\mathrm{SC}}=0$ and $\chi^{\mathrm{SC}}_{\alpha \beta}=\delta_{\alpha \beta} g^{2}\nu_{\mathrm{eff}}^{\mathrm{SC}}/2$. Here, the effective density of states is $\nu_{\mathrm{eff}}^{\mathrm{SC}}=n_{z,\mathrm{max}}m_{\mathrm{SC}}/(2\pi)$. The essential point is that both responses of the SC are isotropic and constant for $B<B_P$: the SC layer provides a large but featureless background, against which the HG signatures discussed below stand out.

\begin{figure}[t]
\includegraphics[width=0.99\columnwidth]{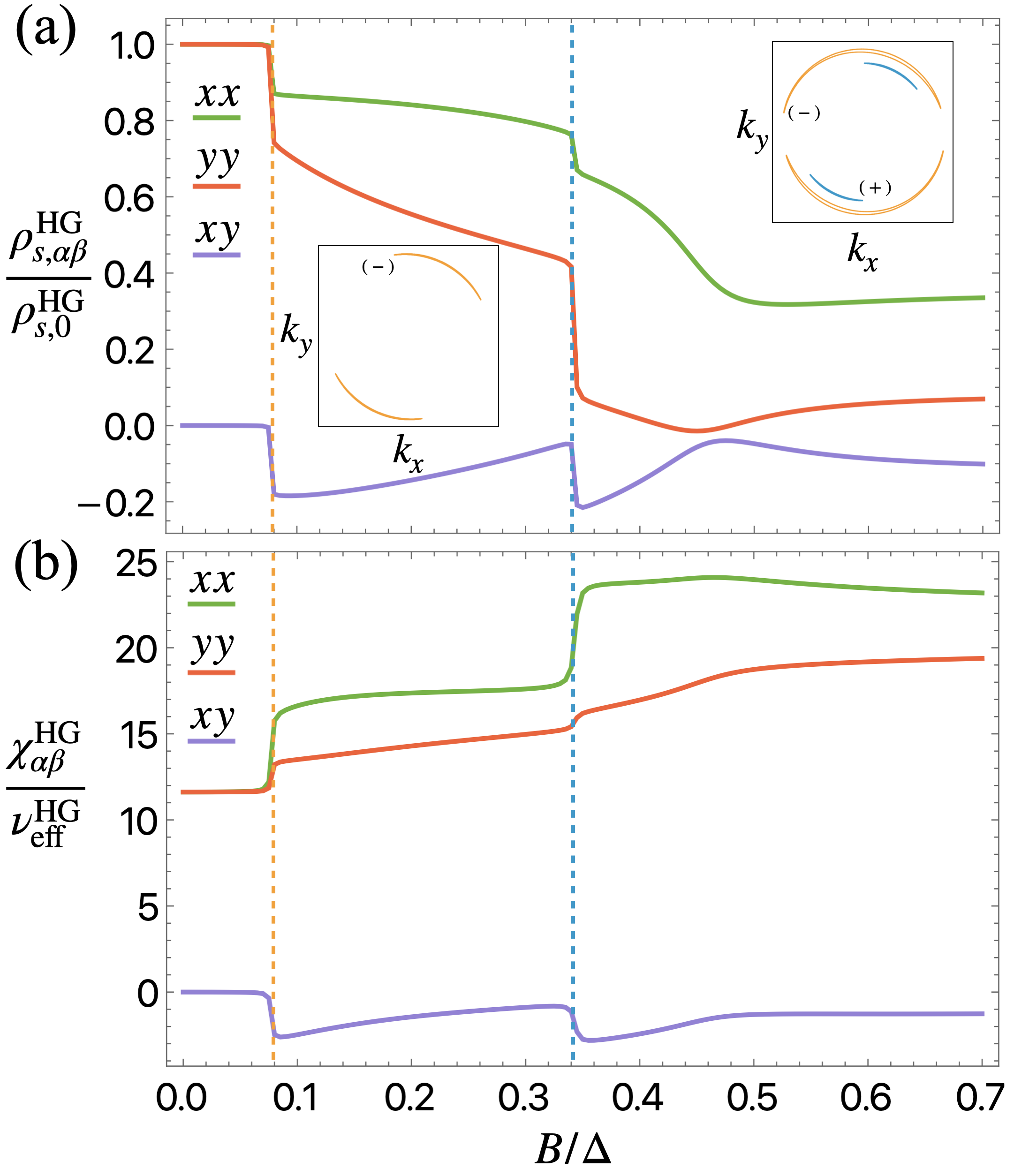}
\caption{\label{fig2} Field-induced non-analyticities in the HG response. Components of (a)~the superfluid-stiffness tensor and (b)~the spin susceptibility as a function of in-plane field $B$ at fixed direction $\phi_B=0$. Steps occur at the gap-closure fields (dashed lines), where a Bogoliubov Fermi surface emerges or changes topology in one of the Rashba-split bands (insets). Note the finite susceptibility $\chi_0$ at $B=0$ and the nonzero $xy$ components, both distinctive of the hybridized HG. The parameters used for these plots are the same as for Fig.\ref{fig1}.}
\end{figure}

\noindent{\it Response of the proximitized HG.---}We now turn to the contributions from the hybridized HG,

\begin{align}
&\rho_{s,\alpha\beta}^{\mathrm{HG}}
=
\rho_{s,0}^{\mathrm{HG}}\intop\frac{d\phi}{2\pi}e_{\alpha}\left(\phi\right)e_{\beta}\left(\phi\right)[2-f(\phi)]
 \label{eq:ns-HG},\\&\chi_{\alpha\beta}^{\mathrm{HG}}=\chi_{0}\delta_{\alpha\beta}+\nu_{\mathrm{eff}}^{\mathrm{HG}}\intop\frac{d\phi}{2\pi}h_{\alpha}\left(\phi\right)h_{\beta}\left(\phi\right) f(\phi) 
\label{eq:chi-HG},\\
& f(\phi)= \sum_{s=
 \pm}\theta([V_{\phi}^{(s)}]^2-[\Delta_{\phi}^{(s)}]^2)\frac{|V_{\phi}^{(s)}|}{\sqrt{[V_{\phi}^{(s)}]^{2}-[\Delta_{\phi}^{(s)}]^{2}}}.
\label{eq:f-func}
\end{align}
We use the vectors $\mathbf{e}=\left(\cos\phi,\sin\phi\right)$ and
$\mathbf{h}=-\left(3\kappa+g\tau_{+-}/2\right)\zeta_{{\rm F}}\mathbf{t}_{\phi}+\left(g\tau_{+-}/2\right)\mathbf{t}_{3\phi-2\phi_{t}}$,
where $\mathbf{t}_{\theta}=(-\sin\theta,\cos\theta)$. The zero-field susceptibility and superfluid stiffness are
$\chi_{0}={\nu_{\mathrm{eff}}^{\mathrm{HG}}}(g^{2}\tau_{+-}^{2}+\zeta_{\rm F}^{2}\left(g\tau_{+-}+6\kappa\right)^{2}+2\zeta_{\rm F}^{2}g^{2}\left|\tau_{z+}\right|^{2})/4$ and $\rho_{s,0}^{\mathrm{HG}}=n_{\mathrm{eff}}^{\mathrm{HG}}/(4|m_{\mathrm{HG}}|)$, where the
effective density of states and carrier density are $\nu_{\mathrm{eff}}^{\mathrm{HG}}=|m_{\mathrm{HG}}|/(2\pi[1+\tau_{+-}])$ and $n_{\mathrm{eff}}^{\mathrm{HG}}=k_{{\rm F},\mathrm{HG}}^{2}/(2\pi[1+\tau_{+-}])$, respectively, while  $\tau_{z+}$ denotes an additional interlayer-tunneling amplitude.

All non-analytic features discussed below stem from the Heaviside-function in $f(\phi)$ in Eq.~(\ref{eq:f-func}), activated when the gap-closure condition $|V_{\phi}^{\pm}|=|\Delta_{\phi}^{\pm}|$ is met. At $B=0$, both tensors are isotropic,
$\rho_{s,\alpha\beta}^{\mathrm{HG}}=\rho_{s,0}^{\mathrm{HG}}\delta_{\alpha\beta}$
and $\chi_{\alpha\beta}^{\mathrm{HG}}=\chi_{0}\delta_{\alpha\beta}$.
The isotropy may seem counter-intuitive since $\Delta_{\phi}^{\pm}$
are anisotropic. However, at $T=0$, a fully gapped superconductor
supports no quasiparticle excitations. As a result, the gap anisotropy
does not manifest itself as long as there is no gap closure. At finite
temperature, thermally excited quasiparticles make the response sensitive
to the gap anisotropy, leading to anisotropic corrections (see Supplemental Material).
Importantly, isotropic does not mean trivial: the finite zero-field
susceptibility $\chi_{0}$ is a distinctive feature of the hybridized
HG. It emerges from the interplay of induced superconducting pairing
and Rashba spin--orbit coupling. 
Physically, it is a Van Vleck-like response of the singlet condensate~\cite{PhysRevB.76.094516}, rather than a Pauli response of broken Cooper pairs: 
$\chi_{0}$ is built from matrix
elements of the Zeeman coupling between the two Rashba-split bands, which are
retained in the magnetic vertex (see Supplemental Material), whereas the corresponding interband terms of the
Hamiltonian are dropped in Eq.~(\ref{H_simp}). This is also why $\tau_{z+}$ contributes to $\chi_{0}$ although it does not appear in Eqs.~(\ref{Eq:xi2x2}--\ref{Eq:Delta2x2}). Specifically, for the calculation of the spin susceptibility, we impose the stronger restriction $3\alpha_{\rm R} k_{\rm F}\zeta_{\rm F} \gg \Delta \tau_{+-}$. This regime is relevant for the fields considered here, $B\lesssim 0.7\Delta$, which must also satisfy the restriction $\kappa B\lesssim\sqrt{\alpha_{\rm R}k_{\rm F}|\xi_{k_{{\rm F}},s}|}$ imposed above. Within this additional restriction, $\chi_0$ is independent of the Rashba strength $\alpha_{\rm R}$ to leading order, results for smaller values of Rashba coupling are presented in the Supplemental Material.

\noindent{\it Field-induced non-analyticities.---}The magnetic-field dependence of $\rho_{s,\alpha\beta}$ and $\chi_{\alpha\beta}$ is shown in Fig.~\ref{fig2}(a-b), respectively. Both exhibit step-like non-analytic features, which originate from the competition between Zeeman energy $V_{\phi}^{\pm}$
and superconducting pairing $\Delta_{\phi}^{\pm}$. The non-analyticities occur when the gap-closing
condition $\min_{\phi}(\Delta_{\phi}^{\pm}\mp V_{\phi}^{\pm})=0$
is reached, at some momentum angle $\phi^{\pm}_0$. Physically, this marks
the onset of Cooper-pair breaking and the emergence of Bogoliubov quasiparticles
with the momentum angle $\phi^{\pm}_0$, which in turn leads to an abrupt
change in superfluid stiffness, since the number of Cooper pairs changes,
and in spin susceptibility, since the emergent quasiparticles carry
a finite effective spin.

We note that the contribution of the hybridized HG to the total superfluid response $\rho_{s,\alpha\beta}=\rho_{s,\alpha\beta}^{\mathrm{SC}}+\rho_{s,\alpha\beta}^{\mathrm{HG}}$ is much smaller than that of the superconducting layer, since the carrier density in the superconductor is much larger than in the HG. However, the superconducting contribution $\rho^{\rm SC}_{s,\alpha\beta}$ is smooth in magnetic field, so the steps of $\rho^{\rm HG}_{s,\alpha\beta}$ [Fig.~\ref{fig2}(a)] are expected to remain visible in the total response. Moreover, the combinations $\rho_{s,xx}-\rho_{s,yy}$ and $\rho_{s,xy}$ have no contribution from the superconducting layer and access the HG directly. The spin susceptibility [Fig.~\ref{fig2}(b)] is even more favorable: the superconducting contribution is suppressed, so the HG features appear directly in the total susceptibility $\chi_{\alpha\beta}^{\mathrm{SC}}+\chi_{\alpha\beta}^{\mathrm{HG}}$.

Now we address the non-analytic features of $\rho_s$ and $\chi$ using the explicit expressions
in Eqs.~(\ref{eq:ns-HG}-\ref{eq:chi-HG}). Fixing the field direction $\phi_{B}$ and sweeping
the field amplitude $|\mathbf{B}|$, as in Fig.~\ref{fig2}, we consider gap closure condition
$\min_{\phi}(\Delta_{\phi}^{\pm}\mp V_{\phi}^{\pm})=0$, which
is reached at some $B=B_{0}^{\pm}$ and $\phi=\phi_{0}^{\pm}$.
Evaluating the angular
integrals for $\rho_{s}$ and $\chi$ in the vicinity
of $\phi_{0}^{\pm}$ yields step-like contributions with heights (see Supplemental Material for details, below we omit $\pm$ index for brevity)
\begin{align}
 & \Delta \rho_{s,\alpha\beta}^{\mathrm{HG}}=-\rho_{s,0}^{\mathrm{HG}}e_{\alpha}(\phi_{0})e_{\beta}(\phi_{0})W_{\phi_{0}}(B_{0})\\
 & \Delta\chi_{\alpha\beta}^{\mathrm{HG}}=\nu_{\mathrm{eff}}^{\mathrm{HG}}h_{\alpha}(\phi_{0})h_{\beta}(\phi_{0})W_{\phi_{0}}(B_{0}),
\end{align}
where $W_{\phi}=|V_{\phi}|/\sqrt{|\partial_{\phi}^{2}[(V_{\phi})^{2}-(\Delta_{\phi})^{2}]|/2}$.
The step heights are thus rank-one tensors along the gap-closing direction: measuring two independent components of a single step in Fig.~\ref{fig2}(a) or (b) fixes $\phi^{\pm}_{0}$ unambiguously.
Similarly, fixing the field amplitude $|\mathbf{B}|$ and sweeping
the angle $\phi_{B}$, we get the gap closure at $\phi_{B}=\phi_{B,0}$
(and generally different $\phi_{0}$). In this case, the step magnitudes take the same form and are evaluated at the corresponding gap-closure points,  see Supplemental Material for details. Thus, the locations and heights of these steps, together with finite $\chi\left(B=0\right)$,
provide direct access to the functions $\Delta_{\phi}^{\pm}$ and
$V_{\phi}^{\pm}$, and therefore constrain microscopic parameters
describing the hybridized HG, specifically, $\tau_{+-}$, $\tau_{z+}$,
$\phi_{t}$, $\zeta_{\rm F}$ and $\zeta_{\rm R}$. Since each step yields three data points (location, height, and gap closing angle) measuring $\rho_{s}$ and $\chi$ as a function of $B$ for a few fixed $\phi_{B}$ (and vice versa) overconstrains these five parameters and offers consistency checks.

\begin{figure}
  \centering
  \includegraphics[width=0.9\columnwidth]{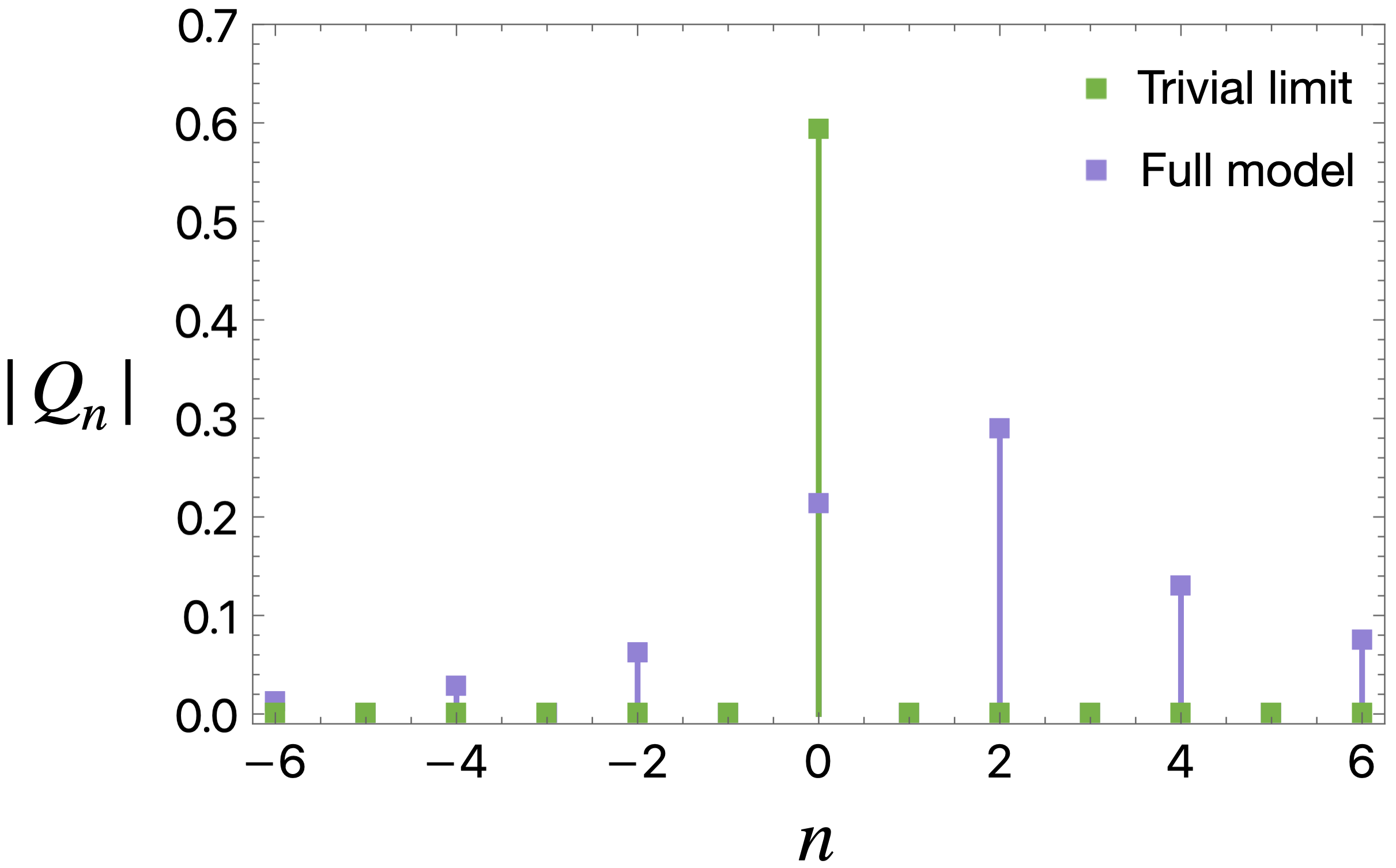}
  \caption{\label{fig3} Fourier harmonics $|Q_n|$ of the angular dependence $Q(\phi_B)$ for $B=0.2 \Delta $ of the superfluid-stiffness tensor, Eq.~(\ref{eq:Q}). In the trivial limit (green), only 
$s$-wave pairing is retained and the cubic-Rashba-related term in
$V_\phi^\pm$ is omitted, yielding only $Q_0\neq0$. In the full model (violet), even harmonics $n=\pm2,\pm4,\pm6$ emerge, which is a direct fingerprint of induced $p$- and $d$-wave pairing and/or cubic Rashba SOC. The parameters used for these plots are the same as for Fig.~\ref{fig1} except $\phi_B$ is varied now.}
\end{figure}

\noindent{\it Angular harmonics of the superfluid stiffness.---}Beyond the step-like features, the angular dependence of the superfluid-stiffness tensor carries qualitative information. We introduce $Q(\phi_B)=e^{-2i\phi_B}\left(\rho_{s,xx}-\rho_{s,yy}+2i\rho_{s,xy}\right)/\rho_{s,0}^{\mathrm{HG}}$, built from the total measurable response: the superconducting layer drops out since $\rho_{s,\alpha\beta}^{\mathrm{SC}}\propto\delta_{\alpha\beta}$. Substituting the expressions for $\rho^{\mathrm{HG}}_{s,\alpha\beta}$ from Eq.~(\ref{eq:ns-HG}), we obtain
\begin{align}
 &
 Q\left(\phi_{B}\right)=\intop\frac{d\phi}{2\pi}e^{2i\left(\phi-\phi_{B}\right)} [2-f(\phi)].
\label{eq:Q}
\end{align}
We then consider the Fourier harmonics of $Q(\phi_B)$: $Q_{n}=\int d\phi_{B}e^{in\phi_{B}}Q\left(\phi_{B}\right)/(2\pi)$. 
It is instructive to first consider the trivial limit, in which only the momentum-independent $s$-wave part of $\Delta_{\phi}^{\pm}$ is retained and the cubic-Rashba-related term in $V_{\phi}^{\pm}$ (the first term) is omitted (shown in green in Fig.~\ref{fig3}). In this case, $Q(\phi_B)$ is independent of $\phi_B$, and therefore only the zeroth harmonic is present: $Q_n\propto\delta_{n,0}$. By contrast, when all terms in the effective Hamiltonian are retained, nontrivial even harmonics with $n\neq 0$ appear (shown in violet in Fig.~\ref{fig3}). In experiment, $Q(\phi_B)$ is accessible by rotating the in-plane field, a standard capability of vector magnets~\cite{Phan_2022}. The emergence of such harmonics thus provides a direct signature of unconventional induced pairing and/or cubic Rashba effects in the hybridized HG.

\noindent{\it Summary and outlook.---}In summary, we calculated superfluid stiffness and spin susceptibility in hybridized HG-SC heterostructures, identifying three signatures of the proximitized HG: field-induced steps, finite $\chi_0$ at $B=0$, and angular harmonics $Q_{n}$. The observables $\chi_{\alpha \beta}$, $\rho_{s,xx}-\rho_{s,yy}$, $\rho_{s,xy}$, and $Q_{n}$ receive no contribution from the superconducting layer and thus isolate the HG from the dominant superconducting background.

These signatures are expected to be robust. Finite temperature and disorder round the steps but do not remove them, as long as the corresponding broadening is small compared to the induced gap $\sim\Delta\tau_{+-}$. The suppression of the parent order parameter $\Delta(B)$ with field shifts the step positions; the harmonics $Q_n$, measured by rotating the field at fixed $|\mathbf{B}|$, are unaffected. Quantitatively, microwave kinetic-inductance measurements resolve sub-percent changes of the superfluid stiffness~\cite{Phan_2022}, while the predicted steps are an $O(1)$ fraction of the HG contribution (Fig.~\ref{fig2}), placing them within experimental reach.

Since the observables studied here are experimentally accessible, our results provide a specific protocol for extracting the intrinsic properties of the hybridized HG. Similar non-analyticities are expected in other thermodynamic quantities, such as the specific heat, and our approach extends directly to other hybrid platforms. We therefore open the way to systematic experimental studies of unconventional proximity-induced phenomena in SC--HG heterostructures.

\begin{acknowledgments}
\noindent{\it Acknowledgments.---}
We are grateful to Georgios Katsaros, Christoph Strunk, Leandro Tosi, and Andrew Higginbotham for useful discussions. 

This research was funded in part by the Austrian Science Fund (FWF) F 86 and in part by the Research Council of Norway through Grant No. 353894. For the purpose of open access, the author/s has/have applied a CC BY public copyright license to any Author Accepted Manuscript version arising from this submission.
\end{acknowledgments}

\bibliography{bib_germanium}

\clearpage
\onecolumngrid
\begin{center}
\textbf{\large Supplemental Material for ``Fingerprints of unconventional pairing in superconductor-hole-gas heterostructures" }\\[5pt]
Serafim S. Babkin,$^{1}$ Karsten Flensberg,$^{2}$ Jeroen Danon,$^{3}$ and Maksym Serbyn$^{1}$\\[5pt]
{\small \sl $^{1}$Institute of Science and Technology Austria (ISTA), Am Campus 1, 3400 Klosterneuburg, Austria}\\
{\small \sl $^{2}$Niels Bohr Institute, University of Copenhagen, DK-2100 Copenhagen, Denmark}\\
{\small \sl $^{3}$Department of Physics, Norwegian University of Science and Technology, NO-7491 Trondheim, Norway}

\vspace{0.1cm}
\begin{quote}
{\small In this Supplemental Material, we present a derivation of
the superfluid stiffness and spin susceptibility using field-theoretical
methods. We start from the partition function derived in our previous
work~\cite{Geproxy} and obtain the effective action by integrating
out the Grassmann fields corresponding to the degrees of freedom in
the superconducting layer (SC) and the hybridized two-dimensional
hole gas (HG). We then derive the superfluid stiffness and spin susceptibility,
which are given by the second derivatives of the action with respect
to the vector potential $\mathbf{A}$ and the Zeeman field $\mathbf{B}$,
respectively.
} \\[10pt]

\end{quote}
\end{center}
\setcounter{equation}{0}
\setcounter{figure}{0}
\setcounter{table}{0}
\setcounter{page}{1}
\setcounter{section}{0}
\setcounter{secnumdepth}{3}
\makeatletter
\renewcommand{\theequation}{S\arabic{equation}}
\renewcommand{\thefigure}{S\arabic{figure}}
\renewcommand{\thesection}{S\arabic{section}}
\renewcommand{\thepage}{\arabic{page}}
\renewcommand{\thetable}{S\arabic{table}}

\vspace{0cm}

\section{Partition function of the heterostructure}

The partition function of the heterostructure is
\begin{align}
 & {\cal Z}=\exp\left[-S\right]=\intop{\cal D}\left[\psi,\overline{\psi},\chi,\overline{\chi}\right]\exp\left[\frac{1}{2}\sum_{n}\left(\sum_{\mathbf{k},n_{z}}\overline{\Psi}_{\mathbf{k}n_{z}n}G_{{\rm SC},n_{z}}^{-1}\Psi_{\mathbf{k}n_{z}n}+\overline{\check{{\cal X}}}_{\mathbf{k}n}G_{\mathrm{HG}}^{-1}\check{{\cal X}}_{\mathbf{k}n}\right)\right],
\end{align}
where $\Psi$ and $\check{{\cal X}}$ are spinors constructed from the Grassmann fields for SC and HG, respectively:
\begin{align}
 & \Psi_{n_{z}}\left(\mathbf{r},\tau\right)=\left[\begin{array}{cccc}
\psi_{\uparrow n_{z}}\left(\mathbf{r},\tau\right) & \psi_{\downarrow n_{z}}\left(\mathbf{r},\tau\right) & \overline{\psi}_{\downarrow n_{z}}\left(\mathbf{r},\tau\right) & -\overline{\psi}_{\uparrow n_{z}}\left(\mathbf{r},\tau\right)\end{array}\right]^{T},\\
 & \check{{\cal X}}_{\mathbf{k}n}=\left(\begin{array}{cccccccc}
\chi_{\mathbf{k}n,3/2} & \chi_{\mathbf{k}n,-3/2} & \overline{\chi}_{-\mathbf{k},-n,-3/2} & -\overline{\chi}_{-\mathbf{k},-n,3/2} & \chi_{\mathbf{k}n,1/2} & \chi_{\mathbf{k}n,-1/2} & \overline{\chi}_{-\mathbf{k},-n,-1/2} & -\overline{\chi}_{-\mathbf{k},-n,1/2}\end{array}\right)^{T}.
\end{align}
Here, $n_{z}$ labels the transverse modes in SC and $G_{{\rm SC},n_{z}}$
is the corresponding Green function:
\begin{align}
 & G_{{\rm SC},n_{z}}^{-1}=i\epsilon_{n}-H_{{\rm SC}},\\
 & H_{{\rm SC}}=\left(\begin{array}{cccc}
\xi_{n_{z}\mathbf{k}} & -\frac{1}{2}gB_{-} & -\Delta\\
-\frac{1}{2}gB_{+} & \xi_{n_{z}\mathbf{k}} &  & -\Delta\\
-\Delta &  & -\xi_{n_{z}\mathbf{k}} & -\frac{1}{2}gB_{-}\\
 & -\Delta & -\frac{1}{2}gB_{+} & -\xi_{n_{z}\mathbf{k}}
\end{array}\right),
\end{align}
where $\epsilon_{n}$ is the Matsubara frequency, $\xi_{n_{z}\mathbf{k}}=\frac{\mathbf{k}^{2}}{2m_{\mathrm{SC}}}-\left[\mu_{s}-\frac{\pi^{2}n_{z}^{2}}{2m_{\mathrm{SC}}d^{2}}\right]\equiv\frac{\mathbf{k}^{2}}{2m_{\mathrm{SC}}}-\mu_{n_{z}}$ is
the dispersion of the $n_{z}$-th transverse mode, $\mu_{s}$ is the
chemical potential of SC, $g$ is the $g$-factor, $m_{\mathrm{SC}}$ is
the effective electron mass, $d$ is the
thickness of the superconducting layer; $B_{\pm}=B_{x}\pm iB_{y}$,
where $B_{x,y}$ are the components of the in-plane magnetic field,
$\Delta$ is the superconducting order parameter.

The Green function $G_{\mathrm{HG}}$ of hybridized HG, which incorporates the interlayer coupling (see \cite{Geproxy} for details), is
\begin{align}
 & G_{\mathrm{HG}}^{-1}=\left(1+\tau_{+-}\right)i\epsilon_{n}-H_{{\rm HG}},\\
 & H_{{\rm HG}}=\left(\begin{array}{cc}
\hat{H}\left(\mathbf{k}\right) & \hat{\Delta}\left(\mathbf{k}\right)\\
\hat{\Delta}^{\dagger}\left(\mathbf{k}\right) & -\sigma_{y}\hat{H}\left(-\mathbf{k}\right)^{T}\sigma_{y}
\end{array}\right),
\end{align}
\begin{align}
 & \Delta_{12}\left(\mathbf{k}\right)=-i\Delta\tau_{+-}\zeta_{R}e^{-i\left(\phi+2\phi_{t}\right)},\\
 & \Delta_{11}\left(\mathbf{k}\right)=\Delta\tau_{+-}+i\sqrt{2}\Delta|\tau_{z+}|\zeta_{B}\sin\left(\phi_{B}-\phi_{t}\right)-\Delta\tau_{+-}\zeta_{F}\cos\left[2\left(\phi-\phi_{t}\right)\right]-\sqrt{2}\Delta|\tau_{z+}|\zeta_{R}\sin\left(\phi-\phi_{t}\right),\\
 & H_{12}\left(\mathbf{k}\right)=-3i\alpha_{{\rm R}}k\zeta_{{\rm F}}e^{-3i\phi}+i\alpha_{{\rm R}}k\tau_{+-}\zeta_{s}e^{-i\left(\phi+2\phi_{t}\right)}-\left(3\kappa+\tfrac{1}{2}g\tau_{+-}\right)B\zeta_{{\rm F}}e^{-i(2\phi+\phi_{B})}+\tfrac{1}{2}g\tau_{+-}Be^{-i\left(\phi_{B}+2\phi_{t}\right)},\\
 & H_{11}\left(\mathbf{k}\right)=H_{22}\left(-\mathbf{k}\right)\Big|_{\mathbf{B}\rightarrow-\mathbf{B}}=\xi_{k}+\xi_{k_{{\rm F}},s}\tau_{+-}\zeta_{{\rm F}}\cos\left[2\left(\phi-\phi_{t}\right)\right]-\left(3\kappa+\tfrac{1}{2}g\tau_{+-}\right)B\zeta_{{\rm R}}\sin\left(\phi-\phi_{B}\right)+\\ \nonumber &+\sqrt{2}\alpha_{{\rm R}}k|\tau_{z+}|\zeta_{s}\sin\left(\phi-\phi_{t}\right)+\sqrt{2}\kappa|\tau_{z+}|B\zeta_{s}\cos\left(\phi_{B}-\phi_{t}\right)+\tfrac{1}{2}\sqrt{2}g|\tau_{z+}|B\zeta_{{\rm F}}\cos\left(2\phi-\phi_{t}-\phi_{B}\right).
\end{align}
Here, the dispersion is $\xi_{k}=k^{2}/2m_{\mathrm{HG}}-\mu$,
where the effective mass is $m_{\mathrm{HG}}=m/(\gamma_{1}+\gamma_{s})$
with $m=-m_{e}$, $\mu$ is the chemical potential, $\gamma_{1},\gamma_{s}$
are dimensionless parameters characterizing the band structure. The dimensionless $\tau_{+-},\tau_{z+},\phi_{t}$ characterize hopping between
SC and HG (see \cite{Geproxy}); $\phi$ and $\phi_{B}$ denote the in-plane
angles of the momentum $\mathbf{k}$ and magnetic field $\mathbf{B}$, respectively;
$\kappa$ is the $g$-factor of the hole gas. Furthermore, $\zeta_{{\rm F}}=\frac{\gamma_{s}k^{2}}{mE_{{\rm HL}}},\zeta_{{\rm R}}=\frac{2\alpha_{{\rm R}}k}{E_{{\rm HL}}},\zeta_{B}=\frac{2\kappa B}{E_{{\rm HL}}},\zeta_{s}=\frac{2\xi_{k_{{\rm F}},s}}{E_{{\rm HL}}}$
are small dimensionless parameters, $\alpha_{R}$ is the Rashba spin-orbit
coupling strength, $E_{{\rm HL}}$ is the energy separation between heavy-hole
and light-hole bands in HG, $\xi_{k_{{\rm F}},s}$ is the kinetic
energy of electrons in the superconductor at the Fermi wave vector
$k_{{\rm F}}$ of the HG.

It is convenient to separate the action into the contribution from the superconductor, \(S_{\mathrm{SC}}\), and the contribution from the hybridized 2D hole gas, \(S_{\mathrm{HG}}\):
\begin{align}
 & {\cal Z}=\exp\left[-S\right]\equiv\exp\left[-S_{\mathrm{SC}}-S_{\mathrm{HG}}\right],\ \ S_{\mathrm{SC}}=-\frac{1}{2}\sum_{n_{z}}\mathrm{Tr}\ln G_{{\rm SC},n_{z}}^{-1},\ \ S_{\mathrm{HG}}=-\frac{1}{2}\mathrm{Tr}\ln G_{\mathrm{HG}}^{-1},
\end{align}
where the expressions for contributions to action are obtained by integrating over the Grassmann fields.
Here and in what follows, $\mathrm{Tr}$ denotes the trace over Nambu
and spin indices, as well as the sum and integration over Matsubara frequency
and momentum, respectively.

\section{Vertices and expansion of the action}

The superfluid stiffness $\rho_{s}$ and spin susceptibility $\chi$ are
obtained from the second derivatives of the action with respect to
$\mathbf{A}$ and $\mathbf{B}$, respectively: 
\begin{align}
 &
\rho_{s,\alpha\beta}=\frac{1}{4}\frac{\partial^{2}S}{\partial A_{\alpha}\partial A_{\beta}},\ \ \chi_{\alpha\beta}=-\frac{\partial^{2}S}{\partial B_{\alpha}\partial B_{\beta}}
.\end{align}
To derive them, we expand the action to second order in small
variations {$\delta\mathbf{A}$ and $\delta\mathbf{B}$}.
We start from the inverse Green function: the vector potential enters
through the shift of momentum and, therefore, generates both a contribution linear in $\delta\mathbf{A}$, denoted by {
${\cal X}_{\mathbf{A}}^{(1)}$}, and a contribution quadratic in {$\delta\mathbf{A}$},
denoted by {${\cal X}_{\mathbf{A}}^{(2)}$},
\begin{align}
 & G^{-1}=G_{0}^{-1}-{\cal X}_{\mathbf{A}}^{(1)}-{\cal X}_{\mathbf{A}}^{(2)},
\end{align}
whereas the Zeeman field $B$ contributes only linearly,
\begin{align}
 & G^{-1}=G_{0}^{-1}-{\cal X}_{\mathbf{B}}^{(1)}.
\end{align}
Here, {$G$ }denotes either the Green function
of the SC or of the hybridized HG, while {$G_{0}$}
is a Green function in the absence of variations
$\delta\mathbf{A}$ and $\delta\mathbf{B}$.

The contribution to the action quadratic in $\delta\mathbf{A}$ is
then $\frac{1}{2}\mathrm{Tr}G_{0}{\cal X}_{\mathbf{A}}^{(2)}+\frac{1}{4}\mathrm{Tr}G_{0}{\cal X}_{\mathbf{A}}^{(1)}G_{0}{\cal X}_{\mathbf{A}}^{(1)}$
while the quadratic in $\delta\mathbf{B}$ contribution is $\frac{1}{4}\mathrm{Tr}\left(G_{0}{\cal X}_{\mathbf{B}}^{(1)}G_{0}{\cal X}_{\mathbf{B}}^{(1)}\right)$. 
For the SC, the vertices are
\begin{align}
 & {\cal X}_{\mathbf{A},\mathrm{SC}}^{(1)}=-\frac{\mathbf{k}\delta\mathbf{A}}{m_{\rm SC}}\tau_{0}\sigma_{0},\ \ {\cal X}_{\mathbf{A},\mathrm{SC}}^{(2)}=\frac{\delta \mathbf{A}^{2}}{2m_{\rm SC}}\tau_{3}\sigma_{0},\ \ {\cal X}_{\mathbf{B},\mathrm{SC}}^{(1)}=-\frac{1}{2}g\tau_{0}\delta\mathbf{B}\boldsymbol{\sigma},
\end{align}
where $\tau_{i}$ and $\sigma_{\alpha}$ are Pauli matrices acting
on Nambu and spin space, respectively. Here, $\delta\mathbf{B}$ has
three components while $\delta\mathbf{A}$ has only two in-plane components,
since orbital motion is confined to the plane.
For hybridized HG, the expressions for vertices
${\cal X}_{\mathbf{A}}$ are as follows,
\begin{align}
 & {\cal X}_{\mathbf{A},\mathrm{HG}}^{(1)}\approx-\frac{\mathbf{k}\delta\mathbf{A}}{m_{\mathrm{HG}}}\tau_{0}\sigma_{0},\ \ {\cal X}_{\mathbf{A},\mathrm{HG}}^{(2)}\approx\frac{\delta\mathbf{A}^{2}}{2m_{\mathrm{HG}}}\tau_{3}\sigma_{0},
\end{align}
here, we retained only the leading contributions to the vertices,
assuming the smallness of the parameters $m_{\mathrm{HG}}/m_{\mathrm{SC}},\zeta_{F}$,
$\zeta_{R},\zeta_{B},\zeta_{s},m_{\mathrm{HG}}\alpha_{R}/k_{F},gB/\xi_{k_{F},s},\Delta/\xi_{k_{F},s}$.

The vertex ${\cal X}_{\mathbf{B},\mathrm{HG}}^{(1)}$ takes the
form,
\begin{align}
 & {\cal X}_{\mathbf{B},\mathrm{HG}}^{(1)}\approx\left(\begin{array}{cc}
{\cal \tilde{X}} & 0\\
0 & -\sigma_{y}\left({\cal \tilde{X}}\Big|_{\mathbf{k}\rightarrow-\mathbf{k}}\right)^{T}\sigma_{y}
\end{array}\right)\\
 & {\cal \tilde{X}}=\left(\begin{array}{cc}
P\left(\phi\right) & R\left(\phi\right)\\
R^{*}\left(\phi\right) & -P\left(\phi\right)
\end{array}\right)\delta B_{z}+\\  \nonumber
 & +\left(\begin{array}{cc}
M\left[\cos\left(2\phi-\phi_{t}\right)\delta B_{x}+\sin\left(2\phi-\phi_{t}\right)\delta B_{y}\right] & L\left(\phi\right)\left(\delta B_{x}-i\delta B_{y}\right)\\
L^{*}\left(\phi\right)\left(\delta B_{x}+i\delta B_{y}\right) & -M\left[\cos\left(2\phi-\phi_{t}\right)\delta B_{x}+\sin\left(2\phi-\phi_{t}\right)\delta B_{y}\right]
\end{array}\right),
\end{align}
where we introduced the notations,
\begin{align}
 & L\left(\phi\right)=-\left(3\kappa+\tfrac{1}{2}g\tau_{+-}\right)\zeta_{{\rm F}}e^{-2i\phi}+\tfrac{1}{2}g\tau_{+-}e^{-2i\phi_{t}},\ \ M=\tfrac{1}{2}\sqrt{2}g|\tau_{z+}|\zeta_{{\rm F}},\\
 & R\left(\phi\right)=\frac{e^{-i\left(2\phi+\phi_{t}\right)}}{\sqrt{2}}g\zeta_{F}\left|\tau_{z+}\right|,\ \ P\left(\phi\right)=-\frac{1}{2}\left(6\kappa+g\tau_{+-}\right)+\frac{1}{2}\zeta_{F}g\tau_{+-}\cos\left[2\left(\phi-\phi_{t}\right)\right],
\end{align}
and also neglected subleading contributions, using the inequalities, $\zeta_{s}\ll\zeta_{F}$, $\zeta_{R}\ll\zeta_{F}$.

\section{Calculation of responses}

We calculate the linear responses using the approximate Green function $\tilde{G}_{\mathrm{HG}}$~\cite{Geproxy}, where Zeeman energy is assumed to be sufficiently small: $\kappa B\lesssim\sqrt{\alpha_{R}k_{\rm F}|\xi_{k_{{\rm F}},s}|}$. To derive $\tilde{G}_{\mathrm{HG}}$, we first rotate
the Green function, $G_{\mathrm{HG}}\rightarrow U^{\dagger}G_{\mathrm{HG}}U$,
where the rotation matrix $U$ is
\begin{align}
 & U=\frac{1}{\sqrt{2}}\left(\begin{array}{cccc}
-ie^{-3i\phi} & 0 & ie^{-3i\phi} & 0\\
1 & 0 & 1 & 0\\
0 & -ie^{-3i\phi} & 0 & ie^{-3i\phi}\\
0 & 1 & 0 & 1
\end{array}\right).
\end{align}
This rotation separates the contributions of the Rashba-split bands
into two diagonal $2\times2$ blocks. Then we neglect the off-diagonal blocks, describing
interband terms,
which yields the block-diagonal Green function $\left(1+\tau_{+-}\right)i\epsilon_{n}-H$,
where the effective Hamiltonian $H$ is defined in the main text.
The approximate Green function is then obtained by rotating back,

\begin{align}
 & \tilde{G}_{\mathrm{HG}}=U\left(\left(1+\tau_{+-}\right)i\epsilon_{n}-H\right)^{-1}U^{\dagger}.
\end{align}

\subsection{Superfluid stiffness}
In this subsection, we present the superfluid stiffness of both the SC and HG. Substituting the simplified Green function into the contribution
to the action quadratic in $\delta\mathbf{A}$, we obtain the superfluid
stiffness for the superconducting layer,
\begin{align}
 & \rho_{s,\alpha\beta}^{\mathrm{SC}}=\delta_{\alpha\beta}\rho_{s,0}^{\mathrm{SC}}\, T\sum_{\epsilon_{n}>0}\Bigg[\frac{i\pi\Delta^{2}}{\left({\left(gB/2-i\epsilon_{n}\right)^{2}-\Delta^{2}}\right)^{3/2}}-\frac{i\pi\Delta^{2}}{\left({\left(gB/2+i\epsilon_{n}\right)^{2}-\Delta^{2}}\right)^{3/2}}\Bigg],
\end{align}
where $\rho_{s,0}^{\mathrm{SC}}=n_{\mathrm{eff}}^{\mathrm{SC}}/(4m_{\mathrm{SC}})$ and $n_{\mathrm{eff}}^{\mathrm{SC}}=k_{F,\mathrm{SC}}^{2}/(2\pi)\sum_{n_{z}=1}^{n_{z,\mathrm{max}}}[1-\pi^{2}n_{z}^{2}/(d^{2}k_{F,\mathrm{SC}}^{2})]$ is an effective carrier density in SC, $m_{\mathrm{SC}}$ is the effective electron mass in the superconductor, and $n_{z,\mathrm{max}}=\lfloor dk_{F,\mathrm{SC}}/\pi\rfloor$
is the number of transverse modes.

The contribution of the hybridized HG to the superfluid stiffness
takes the form
\begin{align}
 & \rho_{s,\alpha\beta}^{\mathrm{HG}}=\rho_{s,0}^{\mathrm{HG}}\left(1+\tau_{+-}\right)T\intop\frac{d\phi}{2\pi}e_{\alpha}e_{\beta}\sum_{\epsilon_{n}>0}\mathcal K\left(\epsilon_{n},\phi\right),
\end{align}
where we introduced the auxiliary function
\begin{align}
 & \mathcal K\left(\epsilon_{n},\phi\right)
 =
\frac{i\pi\left|\Delta_{\mathbf{k}}^{(+)}\right|^{2}}{\left({\left(V_{\mathbf{k}}^{(+)}-i\epsilon_{n}\left(1+\tau_{+-}\right)\right)^{2}-\left(\Delta_{\mathbf{k}}^{(+)}\right)^{2}}\right)^{3/2}}-\frac{i\pi\left|\Delta_{\mathbf{k}}^{(+)}\right|^{2}}{\left({\left(V_{\mathbf{k}}^{(+)}+i\epsilon_{n}\left(1+\tau_{+-}\right)\right)^{2}-\left(\Delta_{\mathbf{k}}^{(+)}\right)^{2}}\right)^{3/2}}+\\
 & +\frac{i\pi\left|\Delta_{\mathbf{k}}^{(-)}\right|^{2}}{\left({\left(V_{\mathbf{k}}^{(-)}-i\epsilon_{n}\left(1+\tau_{+-}\right)\right)^{2}-\left(\Delta_{\mathbf{k}}^{(-)}\right)^{2}}\right)^{3/2}}-\frac{i\pi\left|\Delta_{\mathbf{k}}^{(-)}\right|^{2}}{\left({\left(V_{\mathbf{k}}^{(-)}+i\epsilon_{n}\left(1+\tau_{+-}\right)\right)^{2}-\left(\Delta_{\mathbf{k}}^{(-)}\right)^{2}}\right)^{3/2}},
\end{align} 
and used the following notations:  $\mathbf{e}=\left(\cos\phi,\sin\phi\right)$, $\rho_{s,0}^{\mathrm{HG}}=n_{\mathrm{eff}}^{\mathrm{HG}}/(4|m_{\mathrm{HG}}|)$, with $n_{\mathrm{eff}}^{\mathrm{HG}}=k_{F,\mathrm{HG}}^{2}/(2\pi[1+\tau_{+-}])$.  
Figure~\ref{figsup_ns_chi}(a) shows the magnetic-field dependence of the superfluid-stiffness components at finite temperature. We notice that thermal effects broaden the step-like features demonstrated in the main text. 
\subsection{Spin susceptibility}

\begin{figure*}
\includegraphics[width=\textwidth]{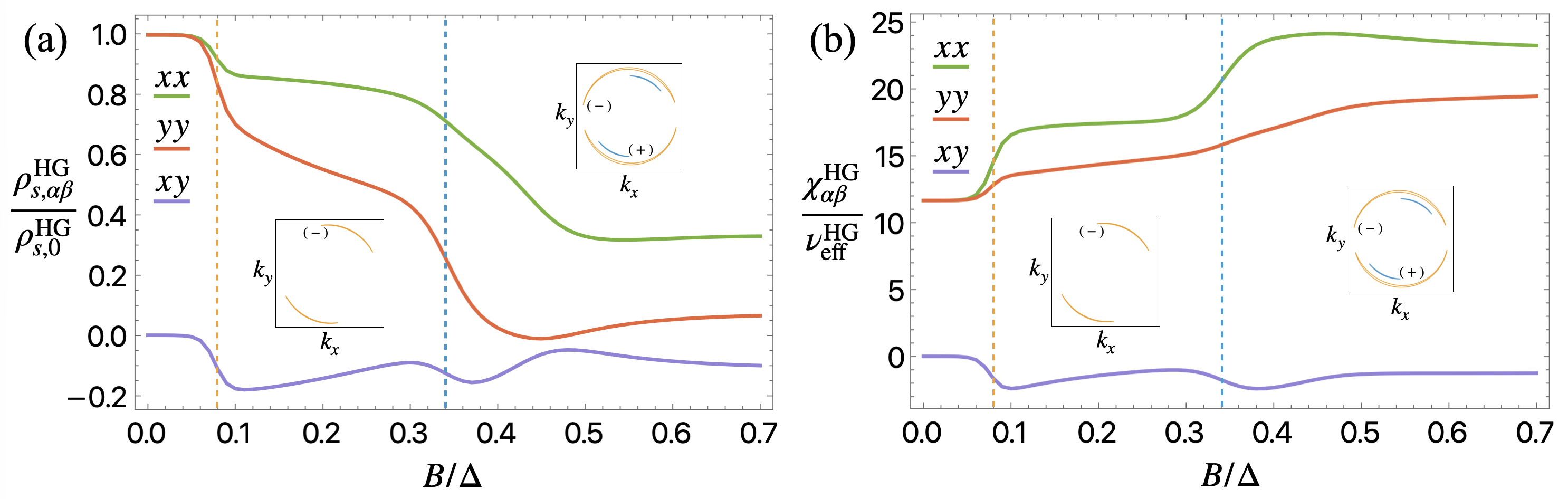}
\caption{\label{figsup_ns_chi} Components of (a) the superfluid-stiffness tensor and (b) the spin-susceptibility tensor as functions of the in-plane field (B) at $\phi_B=0$ and $T=0.02\Delta$. Finite temperature broadens the step-like nonanalyticities.}
\end{figure*}

We now present results for the spin susceptibility.
Since the SC has no intrinsic in-plane anisotropy, we choose the magnetic field along the $x$ axis without loss of generality.
The spin-susceptibility tensor then has only three nonzero components,
$\chi_{||}\equiv\chi_{xx}$, $\chi_{\perp}=\chi_{yy}=\chi_{zz}$,
which are given by
\begin{align}
 & \chi^{\mathrm{SC}}_{||}=n_{z,\mathrm{max}}\frac{m_{\mathrm{SC}}}{2\pi}\frac{g^{2}}{2}\Bigg(1-T\sum_{\epsilon_{n}>0}\Bigg(\frac{i\pi\left|\Delta\right|^{2}}{\left(\sqrt{(gB/2-i\epsilon_{n})^{2}-|\Delta|^{2}}\right)^{3}}-\frac{i\pi\left|\Delta\right|^{2}}{\left(\sqrt{(gB/2+i\epsilon_{n})^{2}-\left|\Delta\right|^{2}}\right)^{3}}\Bigg)\Bigg),\\
 & \chi^{\mathrm{SC}}_{\perp}=n_{z,\mathrm{max}}\frac{m_{\mathrm{SC}}}{2\pi}\frac{g^{2}}{2}\Bigg(1-T\sum_{\epsilon_{n}>0}\Bigg(\frac{gB/2+i\epsilon_{n}}{gB/2}\frac{i\pi}{\sqrt{\left(gB/2+i\epsilon_{n}\right)^{2}-\left|\Delta\right|^{2}}}- 
 \frac{gB/2-i\epsilon_{n}}{gB/2}\frac{\pi i}{\sqrt{\left(gB/2-i\epsilon_{n}\right)^{2}-|\Delta|^{2}}}\Bigg)\Bigg).
\end{align}
For the hybridized HG, the susceptibility tensor takes the form,
\begin{align}
 & \chi^{\mathrm{HG}}_{xx}=\nu_{\mathrm{eff}}^{\mathrm{HG}}\Bigg[\intop\frac{d\phi}{2\pi}\frac{1}{4}\left(e^{-3i\phi}L^{*}\left(\phi\right)-e^{3i\phi}L\left(\phi\right)\right)^{2}\cdot\left(1+\tau_{+-}\right)T\sum_{\epsilon_{n}>0}\mathcal K\left(\epsilon_{n},\phi\right)- \\ \nonumber
 &-\intop\frac{d\phi}{2\pi}\frac{1}{2}\left(\left(e^{-3i\phi}L^{*}\left(\phi\right)+e^{3i\phi}L\left(\phi\right)\right)^{2}+4M^{2}\left(\phi\right)\cos\left(2\phi-\phi_{t}\right)^{2}\right)\cdot\left(1+\tau_{+-}\right)T\sum_{\epsilon_{n}>0}\mathcal{W}\left(\epsilon_{n},\phi\right)+ \\ \nonumber
 &+\frac{1}{2}\left(g^{2}\tau_{+-}^{2}+\zeta_{F}^{2}\left(g\tau_{+-}+6\kappa\right)^{2}+\zeta_{F}^{2}g^{2}\left|\tau_{z+}\right|^{2}\right)\Bigg],\\
 & \chi^{\mathrm{HG}}_{yy}=\nu_{\mathrm{eff}}^{\mathrm{HG}}\Bigg[-\intop\frac{d\phi}{2\pi}\frac{1}{4}\left(e^{-3i\phi}L^{*}\left(\phi\right)+e^{3i\phi}L\left(\phi\right)\right)^{2}\cdot\left(1+\tau_{+-}\right)T\sum_{\epsilon_{n}>0}\mathcal K\left(\epsilon_{n},\phi\right)- \\ \nonumber 
 &-\intop\frac{d\phi}{2\pi}\frac{1}{2}\left(-\left(e^{-3i\phi}L^{*}\left(\phi\right)-e^{3i\phi}L\left(\phi\right)\right)^{2}+4M^{2}\left(\phi\right)\sin^{2}\left(2\phi-\phi_{t}\right)\right)\cdot\left(1+\tau_{+-}\right)T\sum_{\epsilon_{n}>0}\mathcal{W}\left(\epsilon_{n},\phi\right)+ \\ \nonumber
 &+\frac{1}{2}\left(g^{2}\tau_{+-}^{2}+\zeta_{F}^{2}\left(g\tau_{+-}+6\kappa\right)^{2}+\zeta_{F}^{2}g^{2}\left|\tau_{z+}\right|^{2}\right)\Bigg],\\
 & \chi^{\mathrm{HG}}_{xy}=\chi^{\mathrm{HG}}_{yx}=\nu_{\mathrm{eff}}^{\mathrm{HG}}\Bigg[\intop\frac{d\phi}{2\pi}\frac{1}{4}i\left(\left(e^{-3i\phi}L^{*}\left(\phi\right)\right)^{2}-\left(e^{3i\phi}L\left(\phi\right)\right)^{2}\right)\cdot\left(1+\tau_{+-}\right)T\sum_{\epsilon_{n}>0}\mathcal K\left(\epsilon_{n},\phi\right) \\ \nonumber
 &-\intop\frac{d\phi}{2\pi}\frac{1}{2}\left(i\left(\left(e^{-3i\phi}L^{*}\left(\phi\right)\right)^{2}-\left(e^{3i\phi}L\left(\phi\right)\right)^{2}\right)+4M^{2}\left(\phi\right)\cdot\cos\left(2\phi-\phi_{t}\right)\sin\left(2\phi-\phi_{t}\right)\right)\cdot\left(1+\tau_{+-}\right)T\sum_{\epsilon_{n}>0}\mathcal{W}\left(\epsilon_{n},\phi\right)\Bigg],
 \end{align}
\begin{align}
   & \chi_{zz}^{\mathrm{HG}}=\nu_{\mathrm{eff}}^{\mathrm{HG}}\Bigg[\intop\frac{d\phi}{2\pi}\frac{1}{4}\left(e^{-3i\phi}R^{*}\left(\phi\right)-e^{3i\phi}R\left(\phi\right)\right)^{2}\cdot\left(1+\tau_{+-}\right)T\sum_{\epsilon_{n}>0}\mathcal K\left(\epsilon_{n},\phi\right)-\\ \nonumber
 & -\intop\frac{d\phi}{2\pi}\frac{1}{2}\left(4P^{2}\left(\phi\right)+\left(e^{-3i\phi}R^{*}\left(\phi\right)+e^{3i\phi}R\left(\phi\right)\right)^{2}\right)\cdot\left(1+\tau_{+-}\right)T\sum_{\epsilon_{n}>0}\mathcal{W}\left(\epsilon_{n},\phi\right)+ \\ \nonumber
 & +\frac{1}{2}\left(36\kappa^{2}+12g\kappa\tau_{+-}+\frac{1}{2}g^{2}\left(\left(\zeta_{F}^{2}+2\right)\tau_{+-}^{2}+4\zeta_{F}^{2}\left|\tau_{z+}\right|^{2}\right)\right)\Bigg],\\
 & \chi_{xz}^{\mathrm{HG}}=\chi_{zx}^{\mathrm{HG}}=-\nu_{\mathrm{eff}}^{\mathrm{HG}}\Bigg[\frac{1}{4}\intop\frac{d\phi}{2\pi}\left(e^{3i\phi}R\left(\phi\right)-e^{-3i\phi}R^{*}\left(\phi\right)\right)\left(e^{-3i\phi}L^{*}\left(\phi\right)-e^{3i\phi}L\left(\phi\right)\right)\times \\ \nonumber
 & \times \left(1+\tau_{+-}\right)T\sum_{\epsilon_{n}>0}\mathcal K\left(\epsilon_{n},\phi\right)+\frac{1}{2\sqrt{2}}\cos\phi_{t}\zeta_{F}^{2}g\left|\tau_{z+}\right|\left(g\tau_{+-}+12\kappa\right) \\ \nonumber 
 &+\intop\frac{d\phi}{4\pi}\Bigg(\left(e^{-3i\phi}L^{*}\left(\phi\right)+e^{3i\phi}L\left(\phi\right)\right)\left(e^{-3i\phi}R^{*}\left(\phi\right)+e^{3i\phi}R\left(\phi\right)\right)+4P\left(\phi\right)M\left(\phi\right)\cos\left(2\phi-\phi_{t}\right)\Bigg)\left(1+\tau_{+-}\right)T\sum_{\epsilon_{n}>0}\mathcal{W}\left(\epsilon_{n},\phi\right)\Bigg],\\
 & \chi^{\mathrm{HG}}_{yz}=\chi^{\mathrm{HG}}_{zy}=-\nu_{\mathrm{eff}}^{\mathrm{HG}}\cdot\Bigg[\frac{i}{4}\intop\frac{d\phi}{2\pi}\left(e^{3i\phi}R\left(\phi\right)-e^{-3i\phi}R^{*}\left(\phi\right)\right)\left(e^{-3i\phi}L^{*}\left(\phi\right)+e^{3i\phi}L\left(\phi\right)\right)\times \\ \nonumber 
 & \times \left(1+\tau_{+-}\right)T\sum_{\epsilon_{n}>0}\mathcal K\left(\epsilon_{n},\phi\right)+\frac{1}{2\sqrt{2}}\sin\phi_{t}\zeta_{F}^{2}g\left|\tau_{z+}\right|\left(g\tau_{+-}+12\kappa\right) \\ \nonumber
 &+\intop\frac{d\phi}{4\pi}\Bigg(i\left(e^{-3i\phi}L^{*}\left(\phi\right)-e^{3i\phi}L\left(\phi\right)\right)\left(e^{-3i\phi}R^{*}\left(\phi\right)+e^{3i\phi}R\left(\phi\right)\right)+4P\left(\phi\right)M\left(\phi\right)\sin\left(2\phi-\phi_{t}\right)\Bigg)\left(1+\tau_{+-}\right)T\sum_{\epsilon_{n}>0}\mathcal{W}\left(\epsilon_{n},\phi\right)\Bigg].
\end{align}
Here, we introduce the auxiliary function $\mathcal{W}\left(\epsilon_{n},\phi\right)$:
\begin{align}
 & \mathcal{W}\left(\epsilon_{n},\phi\right)=-2\pi\text{Im}\Bigg[\frac{\left(r_{+}-r_{-}\right)\left(r_{+}r_{-}-\left(V_{\mathbf{k}}^{-}-i\epsilon_{n}\left(1+\tau_{+-}\right)\right)\left(V_{\mathbf{k}}^{+}+i\epsilon_{n}\left(1+\tau_{+-}\right)\right)+\Delta_{\mathbf{k}}^{+}\Delta_{\mathbf{k}}^{-}\right)}{r_{+}r_{-}\left(\left(r_{+}-r_{-}\right)^{2}-4A^{2}\right)}\Bigg] \\
 &A=3\alpha_{{\rm R}}k\zeta_{{\rm F}}-\xi_{k_{{\rm F}},s}\tau_{+-}\zeta_{{\rm R}}\cos\left[2\left(\phi-\phi_{t}\right)\right],\ \ r_{\pm}=\sqrt{\left(V_{\mathbf{k}}^{\pm}\pm i\epsilon_{n}\left(1+\tau_{+-}\right)\right)^{2}-\left|\Delta_{\mathbf{k}}^{\pm}\right|^{2}}.
\end{align}
In the regime of strong Rashba $3\alpha_{\rm R} k_{\rm F}\gg \Delta \tau_{+-}$, the contributions containing $\mathcal{W}$ can be neglected, since they are suppressed by big $A$ in denominators. Figure~\ref{figsup_ns_chi}(b) shows the magnetic-field dependence of selected spin-susceptibility components at finite temperature.
\subsection{Step-like features}

In this subsection, we consider the case $T=0$ and analyze the origin of the step-like features in the magnetic-field dependence of the superfluid stiffness and spin
susceptibility. To this end, we introduce the function $G^{(s)}(\phi,B,\phi_{B})=(V_{\phi}^{(s)})^{2}-(\Delta_{\phi}^{(s)})^{2}$
which appears in the denominators of the analytical expressions for
$\rho_{s}$ and $\chi$, where, $s=\pm$ labels the two Rashba-split bands. We first fix the magnetic-field direction $\phi_{B}$ and vary its magnitude $B$. For certain values $B_{0}^{(s)}$,
the local maximum of $G$ reaches zero at $\phi = \phi_{0}^{(s)}$, which corresponds to the gap-closing
condition, $\min_{\phi}(\Delta_{\phi}^{(s)}\mp V_{\phi}^{(s)})=0$,
discussed in the main text. Thus, we can expand near this maximum,

\begin{align}
& G^{(s)}\left(\phi,B,\phi_{B}\right)\approx\partial_{B}G^{(s)}\left(\phi_{0}^{(s)},B_{0}^{(s)},\phi_{B}\right)\left(B-B_{0}^{(s)}\right)+\frac{1}{2}\partial_{\phi}^{2}G^{(s)}\left(\phi_{0}^{(s)},B_{0}^{(s)},\phi_{B}\right)\left(\phi-\phi_{0}^{(s)}\right)^{2}.
\end{align}

Then, the non-analytic contribution to the superfluid stiffness and
spin susceptibility takes the following form,

\begin{align}
 & \intop_{{\cal O}_{\epsilon}\left(\phi_{0}^{(s)}\right)}\frac{d\phi}{2\pi}F\left(\phi\right) f(\phi)
\approx\frac1{2\pi}{F\left(\phi_{0}^{(s)}\right)
\left|V_{\phi_{0}^{(s)}}^{(s)}\right|}\intop_{{\cal O}_{\epsilon}\left(\phi_{0}^{(s)}\right)}d\phi\frac{\Theta\left(G^{(s)}\left(\phi,B,\phi_{B}\right)\right)}{\sqrt{G^{(s)}\left(\phi,B,\phi_{B}\right)}}=\frac{F\left(\phi_{0}^{(s)}\right)\left|V_{\phi_{0}^{(s)}}^{(s)}\right|\Theta\left(B-B_{0}^{(s)}\right)}{\sqrt{2|\partial_{\phi}^{2}G(\phi_{0}^{(s)},B_{0}^{(s)},\phi_{B})|}}.
\end{align}
Here, for the superfluid stiffness $F\left(\phi\right)=-\rho_{s,0}^{\mathrm{HG}}e_{\alpha}\left(\phi\right)e_{\beta}\left(\phi\right)$,
while for the spin susceptibility $F\left(\phi\right)=\nu_{\mathrm{eff}}^{\mathrm{HG}}h_{\alpha}\left(\phi\right)h_{\beta}\left(\phi\right)$, where functions $f(\phi),e_{\alpha}(\phi),h_{\alpha}(\phi)$ are defined in the main text. 

For fixed $B$ and varying field direction $\phi_{B}$, the analogous expansion reads:
\begin{align}
& G^{(s)}\left(\phi,B,\phi_{B}\right)\approx\partial_{\phi_{B}}G^{(s)}\left(\phi_{0}^{(s)},B,\phi_{B,0}^{(s)}\right)\left(\phi_{B}-\phi_{B,0}^{(s)}\right)+\frac{1}{2}\partial_{\phi}^{2}G^{(s)}\left(\phi_{0}^{(s)},B,\phi_{B,0}^{(s)}\right)\left(\phi-\phi_{0}^{(s)}\right)^{2},
\end{align}
then, the integral has the form
\begin{align}
 & \intop_{{\cal O}_{\epsilon}\left(\phi_{0}^{(s)}\right)}\frac{d\phi}{2\pi}F\left(\phi\right) f(\phi)\approx\frac{F(\phi_{0}^{(s)})|V_{\phi_{0}^{(s)}}^{(s)}|}{\sqrt{2|\partial_{\phi}^{2}G(\phi_{0}^{(s)},B,\phi_{B,0}^{(s)})|}} \Theta\left(\partial_{\phi_{B}}G(\phi_{0}^{(s)},B,\phi_{B,0}^{(s)})(\phi_{B}-\phi_{B,0}^{(s)})\right).
\end{align}

Since $G$ is $\pi$-periodic, there is an equivalent local maximum
at $\phi_{0}^{(s)}+\pi$, which gives the same contribution because
$F(\phi)|V_{\phi}^{(s)}|$ is also $\pi$-periodic. Combining these
two contributions, we recover the step heights stated in the main
text.

\subsection{Angular dependence of superfluid-stiffness tensor}

\begin{figure*}[t]
\includegraphics[width=\textwidth]{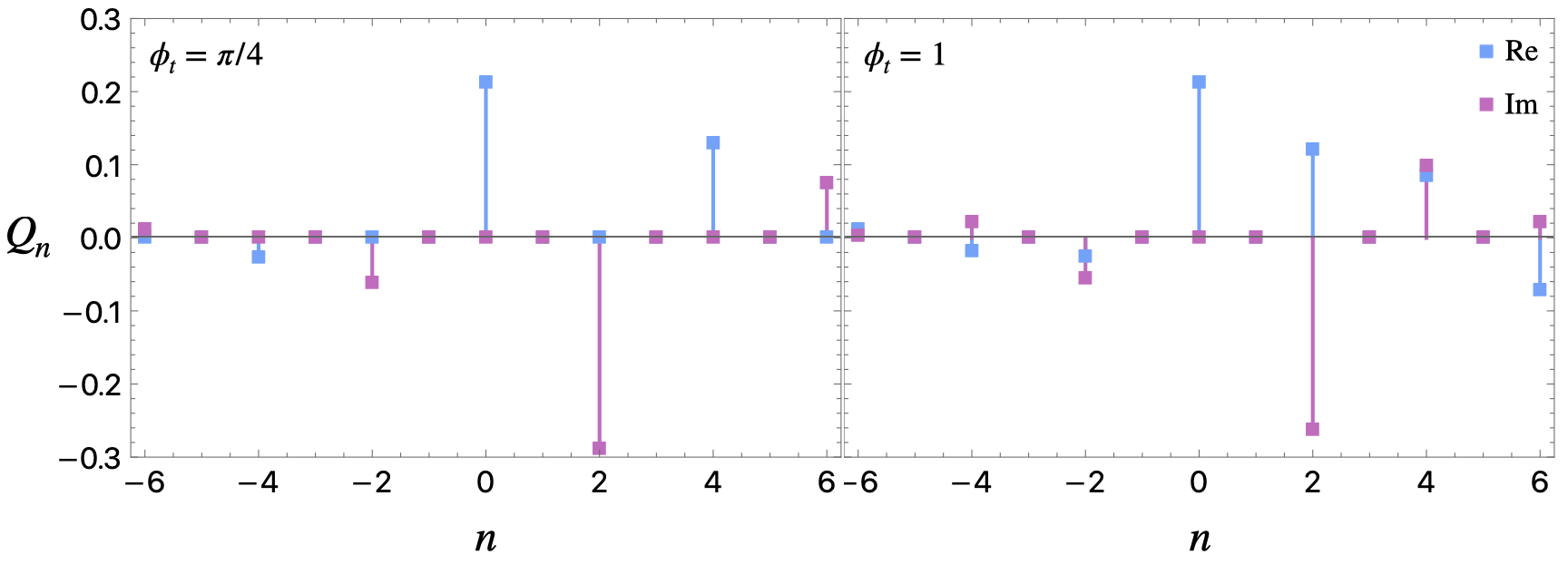}
\caption{\label{figsup_Qn}Fourier harmonics $Q_n$ of the angular dependence $Q(\phi_B)$ of the superfluid-stiffness tensor at $B=0.2\Delta$. The left panel shows the symmetric case, in which the interface is invariant under the exchange $x\leftrightarrow y$, while the right panel shows the asymmetric case.}
\end{figure*}

We define $Q\left(\phi_{B}\right)$ as the
following linear combination of the components of the superfluid stiffness
tensor,

\begin{align}
 & Q\left(\phi_{B}\right)=\frac{e^{-2i\phi_{B}}\left(\rho_{s,xx}-\rho_{s,yy}+2i\rho_{s,xy}\right)}{\rho_{s,0}^{(\mathrm{HG})}}.
\end{align}

Substituting the explicit expressions for the superfluid-stiffness tensor,
we obtain

\begin{align}
 & Q\left(\phi_{B}\right)=\intop\frac{d\phi}{2\pi}e^{2i\left(\phi-\phi_{B}\right)}\left(2-f(\phi)\right).
\end{align}

It is instructive to first consider the trivial limit, in which only
the momentum-independent $s$-wave part of $\Delta_{\phi}^{(\pm)}$
is retained and the cubic-Rashba-related term in $V_{\phi}^{(\pm)}$
(the first term) is omitted. In this case, $f(\phi)$
depends only on $\phi-\phi_{B}$, as a result, $Q(\phi_{B})$ is independent
of $\phi_{B}$, and the Fourier harmonics $Q_{n}=\int d\phi_{B}e^{in\phi_{B}}Q\left(\phi_{B}\right)/(2\pi)$
are non-zero only for $n=0$: $Q_{n}\propto\delta_{n,0}$.

When all terms in the Hamiltonian are retained, additional harmonics
become nonzero, providing a distinct fingerprint of unconventional
pairings and/or cubic Rashba in proximitized HG. Separating the real and imaginary parts, we find that the Fourier
harmonics obey a characteristic selection rule: the real part is nonzero
only for $n=4k$, whereas the imaginary part is nonzero only for $n=2+4k$
(see Fig.~\ref{figsup_Qn}). The absence of odd harmonics follows
from invariance under magnetic-field reversal, $\mathbf{B}\to-\mathbf{B}$,
which implies $Q(\phi_{B})=Q(\phi_{B}+\pi)$. The further separation
between the real and imaginary sectors occurs only when the interface
is invariant under the interchange of the in-plane axes $x$ and $y$,
which implies the constraint on the hopping parameter $\phi_t$: $\phi_t=\pi/4$ and results in the following relations for $Q(\phi_{B})$: $\mathrm{Re}\left[Q\left(\phi_{B}\right)\right]=\mathrm{Re}\left[Q\left(\pi/2-\phi_{B}\right)\right]$,
$\mathrm{Im}\left[Q\left(\phi_{B}\right)\right]=-\mathrm{Im}\left[Q\left(\pi/2-\phi_{B}\right)\right]$.
When this symmetry is broken and $\phi_t\ne \pi/4$, all even harmonics become allowed, as
seen in Fig.~\ref{figsup_Qn}. Thus, the structure of $Q_{n}$
provides a direct probe of the interface symmetry.

\section{Role of disorder and order parameter suppression}

\textit{Disorder.}---In typical devices, the superconducting layer is substantially disordered, whereas the HG remains relatively clean. Accordingly, the clean-limit treatment of the HG used here is experimentally well motivated, while the superconducting layer may require a more refined description for quantitative comparison with measurements. A natural generalization is to model disorder in the superconductor as nonmagnetic impurity scattering and include the corresponding self-energy. This results in a renormalization of the pairing, Zeeman term, and Matsubara frequencies \cite{Babkin_2024}, while the subsequent analysis of the hybridized HG and its response functions proceeds in essentially the same way.

\textit{Order parameter suppression.}---In the main text, we briefly addressed the suppression of the order parameter in the Pauli limit and introduced the critical field $B_P$ at $T=0$. At $T\ne0$, the order parameter acquires a nontrivial field dependence and at sufficiently low temperatures the phase transition is of the first order \cite{Maki1964}. In principle, $\Delta$ is also affected by the coupling to the HG, but this effect is expected to be small because the effective density of states in the superconducting layer is much larger than that in the HG. Thus, to leading order, one can use the standard result for a conventional $s$-wave superconductor in the Pauli limit \cite{Maki1964}. For quantitative fitting of $\rho_s$ and $\chi$ with experimental data, however, it is preferable to use the experimentally measured dependence $\Delta(B)$, as was done in \cite{Phan_2022}.

\end{document}